\documentclass[reprint,twocolumn,superscriptaddress,secnumarabic,amssymb, nobibnotes, aps, pra]{revtex4-1}

\usepackage{amsmath}    
\usepackage{verbatim}    
\usepackage{color}           
\usepackage{subfigure}   
\usepackage{hyperref}    
\usepackage[dvipdfmx]{graphicx}

\usepackage{slashbox}

\begin{document}

\title{DFT+DMFT study on soft moment magnetism and covalent bonding in SrRu$_2$O$_6$}%

\author{Atsushi Hariki}
\affiliation{Institute for Solid State Physics, TU Wien, 1040 Vienna, Austria}
\author{Andreas Hausoel}
\affiliation{Institute for Theoretical Physics and Astrophysics, University of W\"urzburg, Am Hubland 97074 W\"urzburg, Germany}
\author{Giorgio Sangiovanni}
\affiliation{Institute for Theoretical Physics and Astrophysics, University of W\"urzburg, Am Hubland 97074 W\"urzburg, Germany}
\author{Jan Kune\v{s} }
\affiliation{Institute for Solid State Physics, TU Wien, 1040 Vienna, Austria}

\date{\today}

\begin{abstract}
Dynamical mean-field theory is used to study the three orbital model of the $d^3$ compound SrRu$_2$O$_6$ 
both with and without explicitly including the O-$p$ states. Depending on the size of the Hund's coupling $J$,
at low to intermediate temperatures we find solutions corresponding to Mott or correlated covalent insulator. 
The latter can explain the experimentally observed absence of Curie susceptibility in the paramagnetic phase. 
At high temperatures a single phase with smoothly varying properties is observed. SrRu$_2$O$_6$ provides
an ideal system to study the competition between the local moment physics and covalent bonding since
both effects are maximized, while spin-orbit interaction is found to play a minor role.
\end{abstract}

\maketitle

\section{ Introduction}

The 4$d$ and 5$d$ materials with honeycomb lattice have been attracting considerable attention. Na$_2$IrO$_3$ was
proposed to provide a realization of
a topological insulator~\cite{shitade09} or the Kitaev model~\cite{chaloupka10}. 
 Mazin and collaborators pointed out molecular-orbital (MO) features in the electronic structure of Na$_2$IrO$_3$~\cite{mazin12}, 
 which can be also found in materials with similar crystal structure, e.g., $\alpha$-RuCl$_3$, Li$_2$RuO$_3$ or SrRu$_{2}$O$_{6}$.
 The properties of these materials result from competition between Hubbard interaction, Hund's exchange,
 spin-orbit (SO) interaction, band or MO formation, the
relative importance of which depends strongly on the $d$-shell filling.
 
In this article, we use dynamical mean-field theory (DMFT) to investigate the $d^3$ compound SrRu$_{2}$O$_{6}$,
an antiferromagnetic insulator with an unusually high N\'eel 
temperature of 563~K~\cite{hiley15} and a soft magnetic moment. 
A typical insulating magnet is a Mott insulator with 
rigid local moments that survive well above the transition temperature, giving rise to Curie-Weiss susceptibility.
Soft moment insulators are rare, nevertheless, a number of rather different mechanisms exist.
Heisenberg model build from anti-ferromagnetic dimers was studied by some authors~\cite{sachdev90,sommer01}
and several realizations were found~\cite{matsumoto02,kofu09,han06}. 
Excitonic condensation~\cite{kunes15}, proposed in $d^4$~\cite{khaliullin13,jain15} and
$d^6$~\cite{kunes14b,sotnikov16,yamaguchi17}, 
may also lead to soft moment magnetism.
Yet another scenario is provided by correlated covalent insulators~\cite{kunes08,sentef09}.
FeSi is a well known example of a small gap insulator~\cite{schlesinger93,ishizaka05}, which develops a local moment response
at elevated temperature, although it does not order. 
The dimerized (bi-layer) Hubbard model is the simple lattice model exhibiting competition between formation of covalent bonds leading to 
a hybridization gap, on one side, and formation of local moments and Mott gap, on the other side~\cite{castor95}.

The density functional (DFT) calculations for SrRu$_2$O$_6$~\cite{singh15,streltsov15}
with a small gap show a tendency to anti-ferromagnetic 
ordering. The impossibility to stabilize a ferromagnetic solution, on the other hand,
leads the authors of Ref.~\cite{streltsov15} to discard the Mott insulator
local moment picture. This conclusion is supported by the experimental observation of $T$-increasing susceptibility
above $T_N$~\cite{hiley15}. The DFT+DMFT approach was used in Refs.~\cite{streltsov15,okamoto17}.
We comment on these results in section~\ref{sec:4}.

The physics of SrRu$_{2}$O$_{6}$ involves several competing phenomena: covalent bonding and tendency
to form MOs, Hubbard electron-electron repulsion, Hund's coupling and spin-orbit coupling. The authors of 
Ref.~\cite{okamoto17} further pointed out the possibility of an orbital selective Mott state.
Half-filling of the $t_{2g}$ orbitals maximizes Hund's coupling effect and minimizes the interaction strength necessary to get the
Mott state~\cite{mravlje12,luca11}.
The effect of spin-orbit coupling is also suppressed by the half-filling~\cite{ahn17}.
On the other hand, at half filling the honeycomb structure is favorable for the formation of MOs.
This is because for $n=4k+2$, the chemical potential of an (non-interacting) $n$-site cyclic molecule with $n$ electrons falls into the gap.





\label{sec:1}

The paper is organized as follows. 
After introducing the computational method in Sec.~\ref{sec:2}, 
we show the calculated results in Sec.~\ref{sec:3}.
We discuss the results and perform a simple model analysis in Sec.~\ref{sec:4}. 
Finally, Sec.~\ref{sec:5} summarizes our results.

\section{Computational Method}
\label{sec:2}

\begin{figure}
\begin{center}
    \includegraphics[width=85mm]{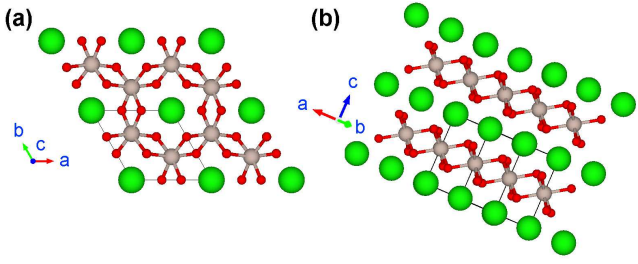}
\end{center}
\vspace{-0.4cm}
\caption{(Color online)
The crystal structure of SrRu$_{2}$O$_{6}$ viewed from two different directions (a) and (b).
The gray, red and green circles represent Ru, O and Sr atoms, respectively.
The crystal structure is visualized using VESTA3 \cite{vesta}}
\label{fig:crys}
\end{figure}

The DFT+DMFT method consists of two steps: (1) construction of an effective model from a converged DFT calculation and
(2) solution of the DMFT self-consistent equation \cite{georges96}.
In this study, we construct two effective models: the $d$-only model (6 bands) describing Ru $t_{2g}$-like bands near the Fermi energy $E_F$ 
and the $dp$-model (24 bands) describing Ru $t_{2g}$ bands + O $2p$ bands.
We perform DFT calculation in the experimental $P{\overline 3}1m$ structure~\cite{hiley14,hiley15}, see~Fig.~\ref{fig:crys},
using the Wien2K package \cite{wien2k} with the generalized gradient approximation (GGA) \cite{perdew96}.
The Wannier functions are constructed 
using the WIEN2WANNIER~\cite{wannier90} and WANNIER90~\cite{wien2wannier} codes.
For the $d$-only case, we construct an effective model including the spin-orbit interaction (SOI)
as well as the one without SOI, for comparison.
\begin{figure}
\begin{center}
 \includegraphics[width=85mm]{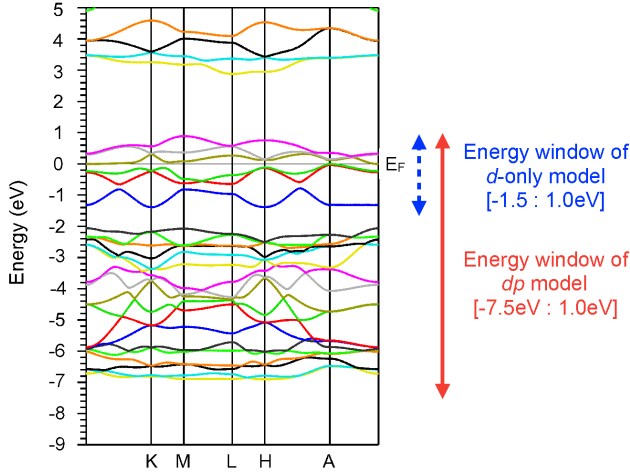}
\end{center}
\vspace{-0.5cm}
\caption{(Color online)
Band structure of SrRu$_{2}$O$_{6}$ with the energy windows 
used in the construction of the $d$-only model and the $dp$ model. 
}
\vspace{-0.4cm}
\label{fig:band}
\end{figure}

The Anderson impurity model (AIM) with the interaction $H_{\rm int}$ between Ru $t_{2g}$ electrons
 \begin{eqnarray}
H_{\rm int}&=&U\sum_{\gamma} n_{\gamma\uparrow}n_{\gamma\downarrow}
       +\sum_{\gamma>\gamma',\sigma}[ (U-2J)n_{\gamma\sigma}n_{\gamma'-\sigma} \nonumber \\
      &+&(U-3J) n_{\gamma\sigma}n_{\gamma'\sigma} ] \nonumber \\
       &+&\frac{\nu}{2}J\sum_{\gamma \neq \gamma',\sigma} 
       (d_{\gamma\sigma}^{\dagger}d_{\gamma' -\sigma}^{\dagger}d_{\gamma\sigma}^{\phantom\dagger} d_{\gamma' -\sigma}^{\phantom\dagger} \nonumber \\
     &-&d_{\gamma-\sigma}^{\dagger}d_{\gamma \sigma}^{\dagger}d_{\gamma'\sigma}^{\phantom\dagger} d_{\gamma' -\sigma}^{\phantom\dagger}), 
\end{eqnarray}
is solved using the
continuous-time
Quantum Monte Carlo method (CT-QMC) in the hybridization expansion formalism \cite{werner06,gull11}.
Here, $\gamma$ and $\sigma$  represent the $t_{2g}$ orbitals and spin, respectively, with
$d_{\gamma\sigma}^{\dagger}$ ($d_{\gamma\sigma}^{\phantom\dagger}$) being the corresponding creation (annihilation) operators
and $n_{\gamma\sigma}$ the number operators.
The calculations for the Slater-Kanamori interaction ($\nu=1$) with spin-rotational symmetry 
were performed with the w2dynamics \cite{parragh12} code. Most of the presented results were obtained with
the density-density approximation ($\nu=0$) for which we used the computationally
efficient segment implementation with
recent improvements \cite{boehnke11,Hafermann12}. 

In the trigonally distorted RuO$_{6}$ octahedra
the $t_{2g}$ levels splits into the $e_{g\pi}$ doublet  and the $a_{1g}$ singlet.
The $e_{g\pi}$ and  $a_{1g}$  crystal-field basis is adopted in the calculation without the SOI.
In the calculation with the SOI, basis diagonalizing the crystal field + SOI
on each Ru atom is adopted.

After self-consistency is achieved, physical quantities of the system 
are obtained from the AIM with the renormalized bath
and the lattice with the converged self-energy $\Sigma(i\omega_n)$.
Besides the static one-particle observables such as occupation number or ordered spin moments we calculated
the following quantities after the self-consistency is reached. The one-particle spectral densities
are obtained by analytic continuation of the self-energy with maximum entropy method \cite{wang09,jarrell96}.
The reduced (diagonal) density matrix providing the weights of atomic states is measured in the QMC simulation
as well as the local spin susceptibility $\chi_{\rm loc}(T)$ given by
\begin{eqnarray}
 \chi_{\rm loc}(T) &=&  \int_0^{1/T} d\tau \langle S_z(\tau)S_z(0) \rangle,
\end{eqnarray}
where $S_z(\tau$) is the spin operator at the imaginary time $\tau$, $T$ is the temperature.
We also show the screen local moment $m_{scr}(T)=\sqrt{ T\chi_{\rm loc}(T)}$.


\section{Results}
\label{sec:3}



\begin{figure}
\begin{center}
   \includegraphics[width=75mm]{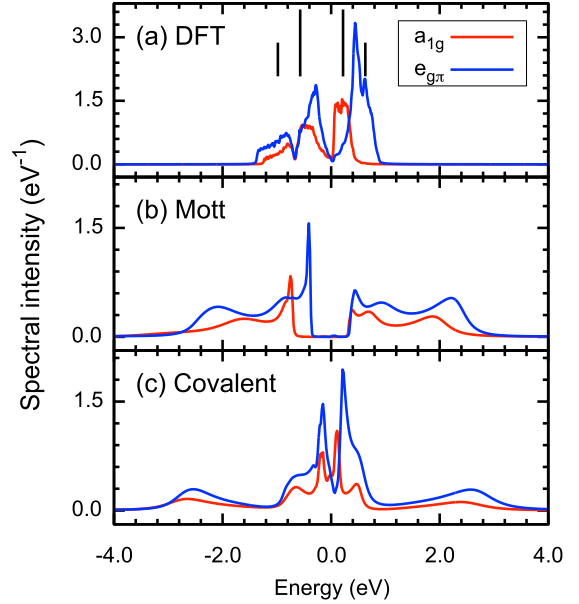}
\end{center}
\vspace{-0.5cm}
\caption{(Color online) Spectral densities of $a_{1g}$ and $e_{g\pi}$ states in the PM phase.
(a) DFT and, (b) Mott and (c) Covalent insulating phase in DFT+DMFT calculation at 500K with $J=0.16$ eV.
The energy origin is take as $E_F$.
The vertical bars in Fig.~(a) represent energy levels of the MOs.} 
\vspace{-0.4cm}
\label{fig:dos}

\end{figure}
In Fig.~\ref{fig:band} we show the DFT band structure with the energy windows
of the $d$-only and the $dp$ models.
The corresponding spectral density with a hybridization gap at the Fermi energy $E_{\rm F}$~ \cite{streltsov15,singh15}
is shown in Fig.~\ref{fig:dos}a. The width of the $e_{g\pi}$ and $a_{1g}$ bands is about $2.2$ eV and 1.8 eV, respectively,
and the orbital occupancies ($d$-only model) are 1.92 and 1.08.
The MO levels~\cite{streltsov15}
of an atomic hexagon with nearest-neighbor hopping amplitude of about 400 meV
are clear in particular in the bonding part of the spectra.
In the DMFT calculations, $U$ is fixed to 2.7~eV and 5.3~eV for the $d$-only  and $dp$ model, respectively,
the values are consistent with random-phase approximation (cRPA) calculations and
previous DMFT studies for Ru compounds \cite{tian15,vaugier12,mravlje11,zhang16}.
Hund's exchange $J$ is treated as an adjustable parameter. In the $dp$ model we use a double-counting 
correction
$\mu_{\rm dc}=(N_{\rm orb}-1)\bar{U}\bar{n}$,
where
$N_{\rm orb}$ is the number of interacting orbitals on a Ru site (6 in our case),
$\bar{U}$ is the averaged Coulomb interaction, and
$\bar{n}$ is the average self-consistent occupation per Ru $d$ orbitals~\cite{krapek12,kunes07}.

\subsection{Paramagnetic phase diagram}
\begin{figure}
\begin{center}
   \includegraphics[width=75mm]{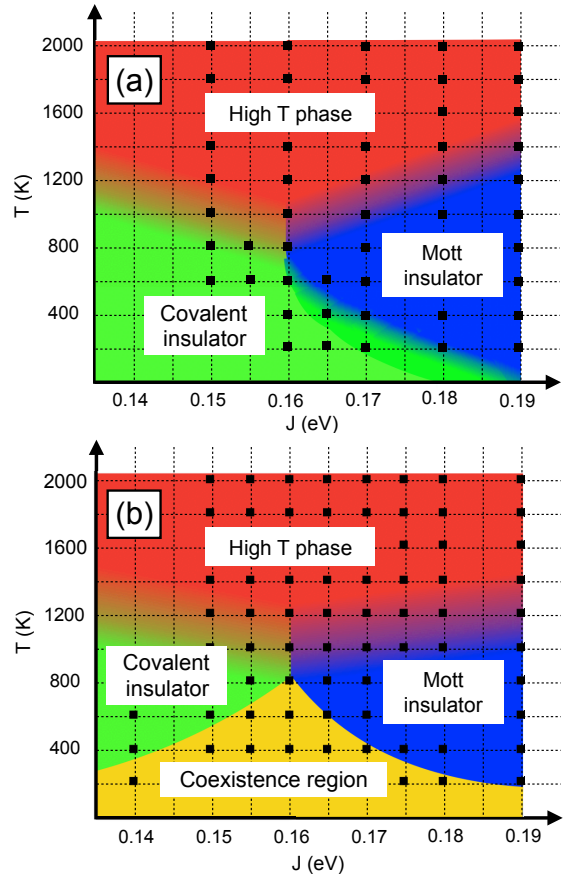}
\end{center}
\vspace{-0.5cm}
\caption{(Color online) 
Phase diagram in the Hund's $J$ and temperature plane 
of the $d$-only model  with (a) Slater-Kanamori interaction and 
(b) density-density interaction.  
The squares mark the points where actual calculations are performed.
Color plot is used for better visibility of different phases.}
\vspace{-0.4cm}
\label{fig:phase}
\end{figure}

First, we present results for the system constrained to the PM phase by symmetrization
over spin in each DMFT iteration.   
Fig.~\ref{fig:phase}b shows the phase diagram in the temperature $T$ vs Hund's value $J$ plane
for the $d$-only model with the density-density interaction.
Below about $T=1000$ K, we obtained two distinct phases: the Mott insulator (MI) and the covalent insulator (CI)~\cite{sentef09,kunes08}.
The MI phase, realized for large $J$, and the CI phase, realized for small $J$, are separated by a coexistence region
where two stable solutions are found.
The MI phase is characterized by presence of fluctuating local moments.
The MI charge gap is opened due to a low-energy peak in the self-energy.
The local moments are absent in the CI phase. The CI charge (pseudo) gap originates from the hybridization.
We quantify these features in the following section.
The charge gap in MI phase of about 0.5~eV, Fig.~\ref{fig:dos}b, is rather large
compared to the experimental gap of about 36~meV~\cite{tian15}.
The pseudo gap is obtained in the CI phase, Fig.~\ref{fig:dos}c,
which appears to match the experiment better.
We note here that the gap in the CI phase is almost independent of $J$.
The coexistence region is likely a non-generic feature.
Its appearance is limited to the $d$-only model with the density-density interaction while it is absent
in the $dp$ model with density-density interaction as well as in the $d$-only model with Slater-Kanamori interaction
in the studied parameters.
%

\begin{figure}


\begin{center}
   \includegraphics[width=75mm]{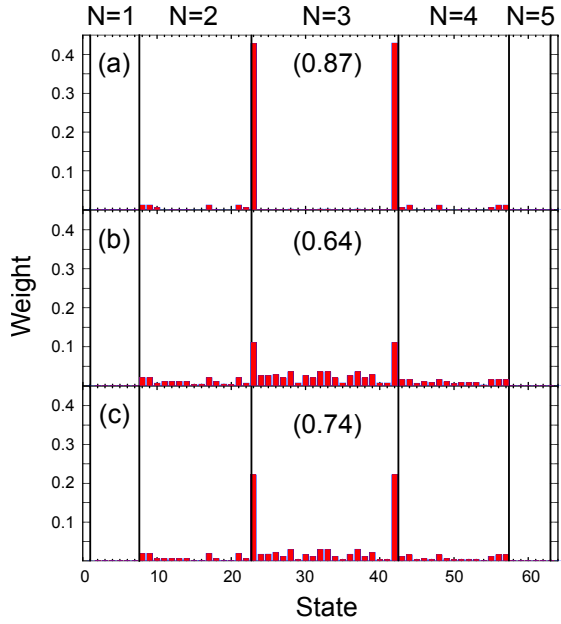}
\end{center}
\vspace{-0.5cm}
\caption{(Color online) The weights of the atomic eigenstates
for (a) Mott insulating phase and (b) covalent insulating phase at 500K, 
and (c) high temperature (1000K).
$J=0.16$ eV is used in the calculation. 
The weight of $N=3$ block is shown in the parentheses.
The two states with the highest weights correspond to $S_z=\pm 3/2$.
}
\label{fig:hist}

\end{figure}

With Slater-Kanamori interaction, Fig.~\ref{fig:phase}a, the hysteretic transition between CI and MI 
is replaced by continuous crossover. Except for the disappearance on the MI phase from most
of the coexistence region the density-density and Slater-Kanamori interactions lead to
similar results. As expected for a $d^3$ system~\cite{ahn17}, the inclusion of SOI has minor 
effect consisting in small shift
(about 0.025 eV)
of the CI/MI crossover to larger $J$.



\begin{figure}
\begin{center}
   \includegraphics[width=90mm]{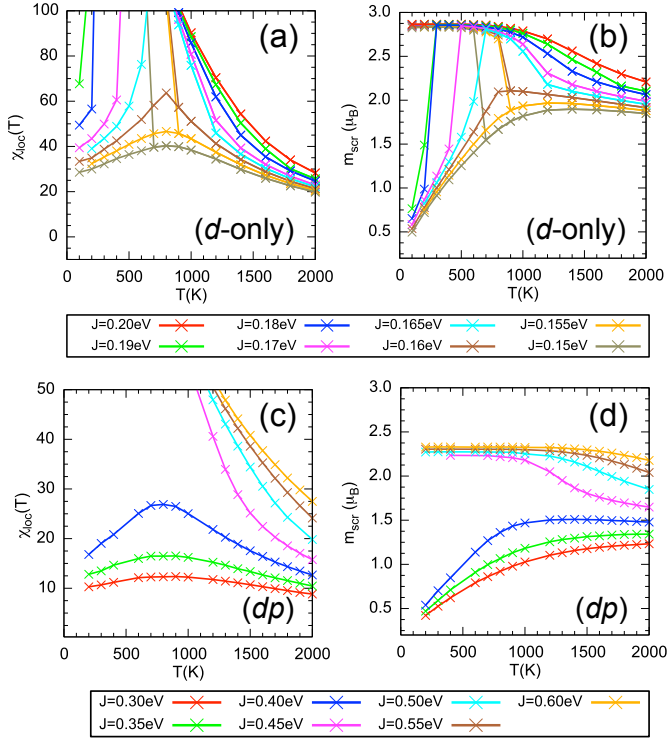}
\end{center}
\vspace{-0.5cm}
\caption{(Color online)
Local susceptibility $\chi_{\rm loc}(T)$ (left) and screened magnetic moment $m_{\rm scr}$ (right)
for the $d$-only model (top) and the $dp$ model (bottom).
In Figs.~(a) and (b), the results for the Mott and the covalent solutions are shown together in the coexistence region.
}
\vspace{-0.4cm}
\label{fig:chi}
\end{figure}

\begin{figure}
\begin{center}
   \includegraphics[width=85mm]{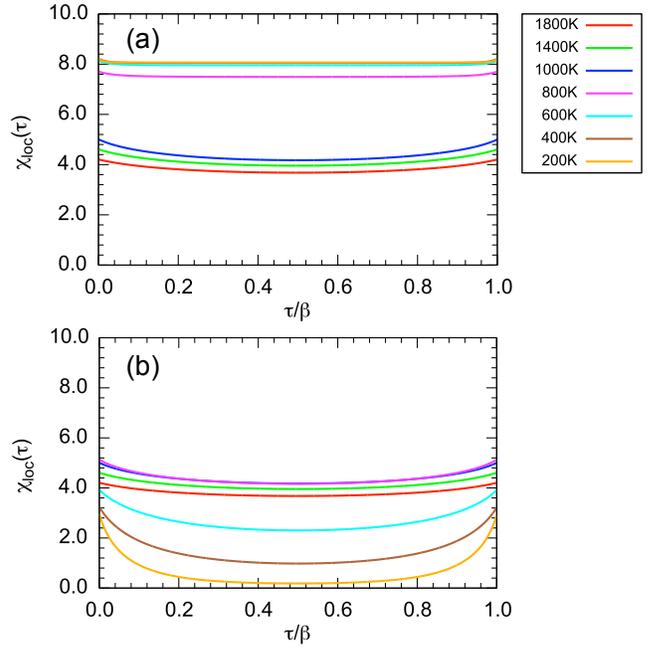}
\end{center}
\vspace{-0.5cm}
\caption{(Color online)
Spin correlation function $\chi(\tau)$ in (a) Mott insulating phase and (b) covalent insulating phase
as a function of $\tau/\beta$, where $\beta=1/k_{\rm B}T$.
The $d$-only model with $J=0.16$ is employed in the calculation.
}
\label{fig:chi_tau}
\end{figure}

\subsection{Charge and spin dynamics in the paramagnetic phase}
\label{sec:3b}

The existence of the coexistence regime allows us to compare the characteristics of CI and MI
states for the same parameters. The atomic state weights 
in the MI and CI phases, characteristic for these phases in general,
are shown in Fig.~{\ref{fig:hist}a,b.
These are the diagonal elements of the site-reduced density matrix, which measure 
the relative time spent by the system in a given atomic state. 
The MI phase is dominated by the high-spin $S=3/2$ states within the 
$d^3$ sector.  The CI phase exhibits larger charge fluctuations -- more weight
in the $d^2$ and $d^4$ sectors -- and much more even population
of the $d^3$ states. These differences reflect the formation of atomic high-spin states
in MI and MOs in CI. Unsurprisingly, the high temperature weights are somewhere
between the two.

Next, we show the local susceptibility $\chi_{\rm loc}(T)$ obtained with the density-density interaction.
First, $\chi_{\rm loc}(T)$ of the $d$-only model is shown in
Fig.~\ref{fig:chi}a. 
In the MI phase, the $\chi_{\rm loc}(T)$ shows the Curie $1/T$ behavior 
with a local moment $m_{\rm scr}$ close to the atomic value of 3~$\mu_B$. 
At high temperatures, $m_{\rm scr}$ is somewhat reduced due to admixture 
of other atomic states, see Fig.~\ref{fig:chi}b.
In the CI phase, on the other hand, the $\chi_{\rm loc}(T)$ is linearly increasing with $T$ 
until it reaches the maximum around $800$K and starts to decrease.
The peak position in $\chi_{\rm loc}(T)$ is almost independent of $J$, but
correlates with the size of the (non-interacting) hybridization gap,
as discussed in Sec.~\ref{sec:4}. 
The $m_{\rm scr}$ in the CI phase does not have much of a physical meaning.

The behavior of $\chi_{\rm loc}$ is determined by the local spin-spin
correlation $\chi_{\rm loc}(\tau) = \langle S_z(\tau)S_z(0) \rangle$,
shown in Fig.~\ref{fig:chi_tau} for $J=0.16$. 
The $\chi_{\rm loc}(\tau)$ in the MI phase shows a typical 
$\tau$-constant behavior. In the CI phase, the $\chi_{\rm loc}(\tau)$ 
rapidly decays from relatively large instantaneous values of $\langle S_z^2 \rangle$,
reflecting the sizeable weights of the high-spin states in Fig.~{\ref{fig:hist}b.
This shows that $\langle S_z^2 \rangle$ itself is not enough
to draw conclusions about the nature of magnetic response.

%
The $\chi_{\rm loc}(T)$ of the $dp$ model is shown in
Fig.~\ref{fig:chi}c. With rescaled parameters it exhibits similar behavior as in the $d$-only model.
Especially, the $\chi_{\rm loc}(T)$ for $J=0.40$ eV gives a peak around $800$ K similar to
the one of the $d$-only model at $J=0.16$ eV. In the $dp$ model,
the MI phase appears for $J>0.45$ eV.
The low-$T$ $m_{\rm scr}$ is only about 2.3 $\mu_{B}$ (Fig.~\ref{fig:chi}d), reflecting
the different Wannier basis and explicit presence of the O-$p$ states, shown in 
Fig.~\ref{fig:wannier}.

\begin{figure}
\begin{center}
   \includegraphics[width=50mm]{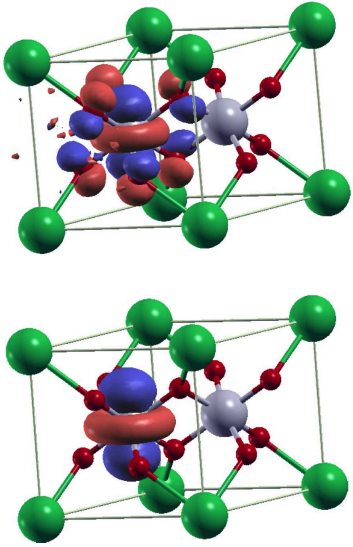}
\end{center}
\vspace{-0.5cm}
\caption{(Color online)
The $a_{1g}$ wannier function in the $d$-only model (top) and $dp$ model (bottom). 
The gray, red and green circles represent Ru, O and Sr atoms, respectively.
The wannier function in the $d$-only model has a considerable weight on neighboring O sites.
}
\vspace{-0.0cm}
\label{fig:wannier}
\end{figure}

\subsection{Anti-ferromagnetic phase}
When the paramagnetic constraint is lifted the system picks the G-type AF order at lower temperatures.
Figs.~\ref{fig:af}a,c show the ordered (staggered) moment per atom in the AF phase.
Both the $d$-only and $dp$ calculations overestimate the N\'eel temperature $T_{\rm N}$ substantially. The Slater-Kanamori interaction
reduces $T_{\rm N}$ by about 200~K, which still leaves a substantial overestimation. This is, however, not unexpected.
The lack of non-local correlations~ \cite{rohringer11} in our approach, particularly in a material with layered (quasi-2D) structure, can be blamed. 

The ordered moment also appears to be substantially larger than the experimentally reported value of 1.3~$\mu_B$.
The two quantities are, however, not directly comparable. Unlike uniform magnetization which is unique, the staggered moment corresponds to 
and depends on overlapping Wannier orbitals~\cite{mravlje12}. In particular, the more delocalized orbitals of the $d$-only model
may be quite bad for comparison with experimental moments. Therefore we calculate the ordered moments  in non-overlapping
atomic spheres. 
The average spin magnetic moment in the muffin-tin sphere ($R_{\rm MT}=1.96$ a.u.) at site $l$ is given by
\begin{equation}
\begin{split}
&\mathbf{m}_l=\sum_{\alpha,\beta,n\gamma,n'\gamma'} \boldsymbol{\sigma}_{\alpha\beta}
\langle d^{\dagger}_{n\gamma\alpha}
d^{\phantom\dagger}_{n'\gamma'\beta} \rangle 
\rho^l_{n\gamma,n'\gamma'}\\
&\rho^l_{n\gamma, n'\gamma'}=\langle w_{n'\gamma'}|P^l_{\text{MT}}|w_{n\gamma}\rangle,
\end{split}
\end{equation}
where $w_{n\gamma}$ is the Wannier function (WF) on lattice site $n$ carrying the orbital flavor $\gamma$ and
$P^l_{\text{MT}}$ is the projection operator on the muffin-tin sphere on site $l$. Selected values of
$\rho^l_{n\gamma,n'\gamma'}$ are shown in Table~\ref{tab:overlap}.
The dominant contribution 
to the reduction of $\mathbf{m}_l$ from its value per Wannier orbital comes from the $l=n=n'$ terms, i.e. the leakage 
of WFs from the muffin-tin spheres.
The nearest-neighbor terms  with $l\neq n=n'$ cause correction of about 0.5\%.
The contribution from the overlap terms $l=n\neq n'$ is negligibly small in the $dp$ model and is precisely zero in the $d$-only model.
These results show that DFT+DMFT yields ordered moments of the same size as the LDA~\cite{streltsov15,singh15} and that
the size of the ordered moment cannot be used as a criterion to distinguish between the Mott and covalent insulator scenarios.

\begin{figure}
\begin{center}
   \includegraphics[width=85mm]{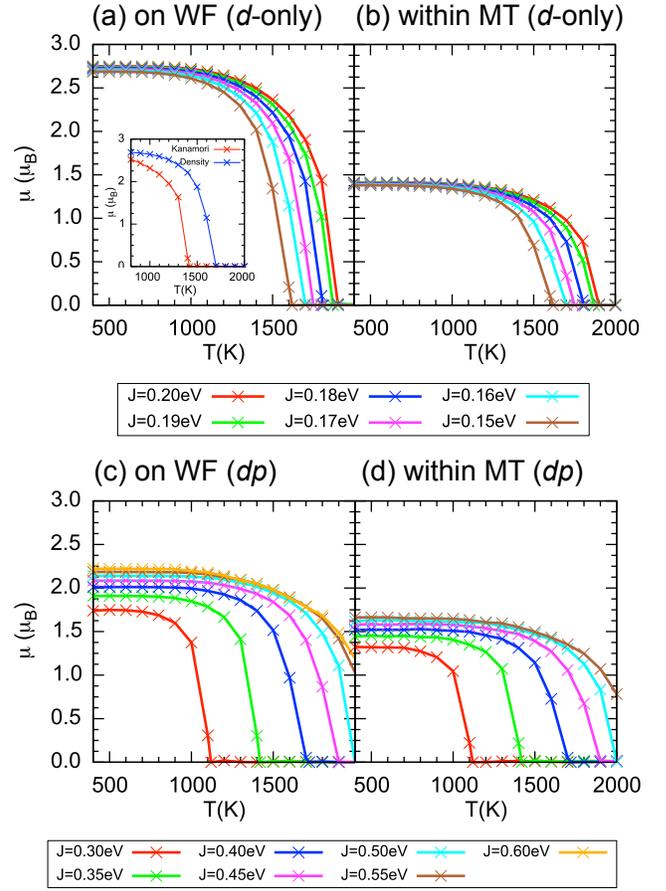}
\end{center}
\vspace{-0.5cm}
\caption{(Color online)
Temperature dependence of the magnetic moment $\mu$ for the $d$-only model (top) and the $dp$ model (bottom), respectively.
The $\mu$ computed on wannier functions (WF) and within muffin-tin (MT) sphere are compared. 
The density-density interaction is employed in the calculation.
The inset in (a) shows a comparison of the (full) Slater-Kanamori and density-density calculations with $J=0.16$.}
\vspace{-0.4cm}
\label{fig:af}
\end{figure}

{\renewcommand\arraystretch{1.6}
\begin{table}
%
\vspace{+0.0cm}
\scalebox{1.2}[1.1]{
\begin{tabular}{|c|c|c|c|c|}
\multicolumn{4}{l}{(a) $\rho^n_{n \gamma,n \gamma'}$} \\[3pt]
\hline
\backslashbox{$\gamma$}{$\gamma$'} & $a_{1g}$ & $e^{(1)}_{g\pi}$ & $e^{(2)}_{g\pi}$ \\\hline
$a_{1g}$ & \ 0.545 \ & \ 0.000 \ & \ 0.000 \ \\ \hline
$e^{(1)}_{g\pi}$  &   0.000  &   0.518  &  0.000  \\ \hline
$e^{(2)}_{g\pi}$  &0.000   & 0.000  &    0.518 \\ \hline 
\end{tabular}
}
\scalebox{1.2}[1.1]{
\begin{tabular}{|c|c|c|c|c|}
\multicolumn{4}{l}{(b) $\rho^l_{n \gamma,n \gamma'}$  ($n$ $\in$ $n.n$ of $l$ in $ab$ plane) } \\[3pt]
\hline
\backslashbox{$\gamma$}{$\gamma$'} & $a_{1g}$ & $e^{(1)}_{g\pi}$ & $e^{(2)}_{g\pi}$ \\\hline
$a_{1g}$ & \ 0.007 \ & \ -0.006 \ & \  0.000 \ \\\hline
$e^{(1)}_{g\pi}$  &   -0.006  &   0.005  &  0.000  \\\hline
$e^{(2)}_{g\pi}$  &  0.000  & 0.000  &    0.002 \\\hline
\end{tabular}
}
\caption{Computed values of the projection operator $\rho^l$ on the muffin-tin sphere $l$ for the $d$-only model
(see Eq~(3)).}
  \label{tab:overlap}
\end{table}


\section{Discussion and model analysis}
\label{sec:4}
Previously, SrRu$_{2}$O$_{6}$ has been studied with several experimental and theoretical methods.
The basic experimental features of SrRu$_{2}$O$_{6}$ are insulating behavior, AF order and the lack of 
local moment response above $T_N$. The DFT calculations~\cite{singh15,streltsov15} found a small hybridization gap and tendency to the AF ordering.
Moreover, the fact that a ferromagnetic DFT solution
cannot be stabilized~\cite{singh15,streltsov15} indicates that the material is not a local moment
magnet. Streltsov {\it et al.}~\cite{streltsov15} performed a basic DFT+DMFT calculation and obtained a local moment Mott insulator --
consistent with our results for large $J$. Okamoto {\it et al.}~\cite{okamoto17} used DFT+DMFT and the $dp$ model.
Varying the double-counting correction
they arrived at an orbitally selective Mott state, which according to the authors provides the best description
of SrRu$_{2}$O$_{6}$. Our results lead to the correlated covalent insulator scenario, which is qualitatively different. Given the
available data, we do not think it is possible to decide which of the two scenarios is more realistic.
Since we did not find the selective
Mott state in our calculations we conclude
that the main difference between our model and Ref.~\onlinecite{okamoto17} is the form of the interaction.
While the present interaction does not distinguish between the $t_{2g}$ orbitals, the authors of Ref.~\onlinecite{okamoto17} use
about
5\%
stronger repulsion between the $a_{1g}$ electrons than between the $e_{g\pi}$ ones.

The correlated insulator picture provides a natural explanation for the lack of local moment behavior. 
The overall shape of $\chi_{\rm loc}(T)$ in the 3-orbital is the same as in MO one-orbital model~\cite{kunes08} and follows
from the competition between formation of local moments on one side and bonding/anti-bonding orbitals on the other.
The multi-orbital nature of the atoms and Hund's coupling $J$ favor the local moment formation. Nevertheless,
numerical data show that the position of the $\chi_{\rm loc}(T)$  maximum is rather insensitive to $J$.
Model DMFT calculations in  Fig.~\ref{fig:model} show that the maximum shifts when the hybridization gap is changed.

\begin{figure}
\begin{center}
   \includegraphics[width=85mm]{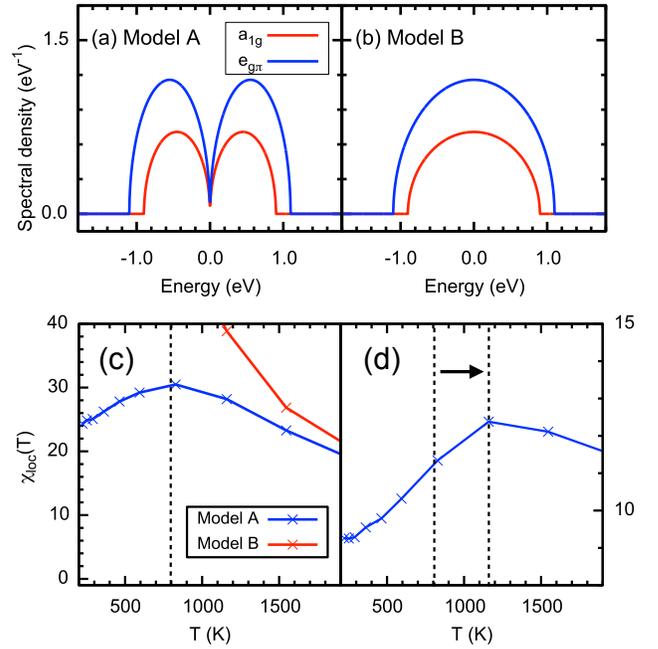}
\end{center}
\vspace{-0.5cm}
\caption{(Color online)
The spectral densities  of (a) model A with a tiny gap at $E_F$ and (b) model B with simple semiellipses.
(c) $\chi_{\rm loc}(T)$ in model A and B. 
Model A gives a similar $\chi_{\rm loc}(T)$ of CI in realistic model (Fig.~\ref{fig:chi}a,c), while model B shows a Curie $1/T$ behavior.
(d) {\it hybridization gap dependence:} $\chi_{\rm loc}(T)$ in model A with a large hybridization gap (0.1eV).
The peak is shifted about 400 K, as indicated by an arrow. 
Here, $U=2.7$ and $J=0.14$ eV  is commonly employed in the DMFT calculation.}
\vspace{-0.4cm}
\label{fig:model}
\end{figure}

\begin{figure}
\vspace{+0.5cm}
\begin{center}
   \includegraphics[width=0.45\columnwidth]{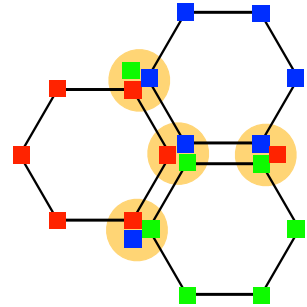}
\end{center}
\vspace{-0.5cm}
\caption{(Color online)
The model of coupled hexagons following Mazin {\it et al.}~\cite{mazin12}. Each of the $t_{2g}$
orbitals belongs to one of the three hexagons that meet at the Ru atom. The hexagons
are formed by the bonds of maximal hopping, inter-hexagon hopping is neglected. The hexagons
are coupled by inter-orbital Coulomb interaction on each atom.}
\vspace{-0.4cm}
\label{fig:cartoon}
\end{figure}

While our results show that DMFT is able to capture the competition between formation 
of hybridization gap and local moments, the question remains how accurate such a description is
in systems consisting of molecular building blocks such as dimers or hexagons. 
If the inter-layer hopping in bi-layer models is smaller or comparable to the intra-layer one~\cite{kunes08,sentef09},
the DMFT description should not be worse than it is for a single-layer system.
However, when the inter-layer hopping substantially exceeds the intra-layer one, DMFT with a
'dimer site' is a more natural description.
The 'atom' and 'dimer' DMFT may lead to rather different
results~\cite{biermann05}. The discussed DMFT studies of SrRu$_{2}$O$_{6}$ assume that 
the non-local correlations within the Ru-hexagons can be neglected.
An opposite extreme
would be a model in Fig.~\ref{fig:cartoon} of isolated hexagons (neglecting the inter-hexagon hopping) 
carrying six electrons each, coupled by Coulomb interaction at the vertices. Comparing
the magnetic properties of such model with the present DMFT may provide further 
insight in the physics of SrRu$_{2}$O$_{6}$ and materials with quasi-molecular structure in general.

\section{Conclusion}
\label{sec:5}
Using DFT+DMFT approach we have studied 3-orbital $d$-only and $dp$ models of SrRu$_{2}$O$_{6}$. 
Depending on the value of the Hund's exchange $J$ we find at temperatures below approximately
800-1000~K either Mott insulator with local moments (large $J$) or a covalent insulator (small $J$). 
Above 1000~K a single regime with properties smoothly varying with $J$ is observed. 
Comparing to the experimental observations we conclude that the covalent insulator regime with
$T$-increasing local susceptibility is realized in SrRu$_{2}$O$_{6}$. We point out that an alternative
scenario based on orbitally selective Mott physics was proposed in Ref.~\onlinecite{okamoto17}.
We find strong tendency to anti-ferromagnetic order for all studied values of the Hund's exchange $J$ 
with $T_N$ substantially exceeding the experimental value, 
which we attribute to layered crystal structure and the mean-field nature of our theory. 
The role of possibly strong non-local correlation reflecting the molecular orbital nature of the compound
remains an open question.

\begin{acknowledgments}
The authors thank V. Pokorn\'y,  A. Sotnikov and J. Fernandez Afonso for fruitful discussions.
This work has received funding from the European Research Council (ERC)
under the European Union's Horizon 2020 research and innovation programme (grant agreement No.646807-EXMAG).
\end{acknowledgments}

\bibliography{SrRu2O6}

\end{document}